# Optically induced spin-dependent diffusive transport in the presence of spin-orbit interaction for all-optical magnetization reversal

Mehrdad Elyasi and Hyunsoo Yang[*]

Department of Electrical and Computer Engineering, National University of Singapore, 117576 Singapore

We have considered the effect of different spin-orbit interaction mechanisms on the process of demagnetization under the influence of short-pulse lasers. All-optical magnetization reversal of perpendicularly magnetized thin films can occur if there are sufficient strong spin-Hall, skew scattering, and Rashba interactions. In the presence of spin-orbit interactions, the transient charge currents provide the generation of transverse-spin currents and accumulations, which eventually exert spin-transfer torque on the magnetization. By combining the optically excited spin-dependent diffusive transport with the spin and charge currents due to skew scattering, spin-Hall, inverse spin-Hall and Rashba interactions, into a numerical model, we demonstrate a possibility of ultra-fast all-optical magnetization reversal. This understanding provokes intriguing, more in-depth experimental studies on the role of spin-orbit interaction mechanisms in optimizing structures for all-optical magnetization reversal.

[*]eleyang@nus.edu.sg

# I. INTRODUCTION

Optical manipulation of spins has been intensively studied for different applications. Coherent coupling of light with spin dynamics can be explained through relativistic quantum electrodynamics, and the effect is pronounced as a reservoir of angular momentum in spin manipulation of structures such as antiferromagnetic NiO or solid-state defects.[1-4] However, the effect of light on the spin dynamics is applicable to not only coherent couplings, but also optical spin-transfer torque, modification of crystalline anisotropy, and demagnetization.[5-23] The origin of the demagnetization in different structures itself has been subject to an ongoing debate.[8,9,11,14-18,20,24-28] The Elliot-Yafet mechanisms based on electron-phonon and electron-electron interactions, as well as magnonic dissipation and spin-dependent super-diffusive transport have been discussed to explain the ultra-fast demagnetization.

Incoherent all-optical manipulation of spin and magnetization reversal in ferrimagnets, anti-ferromagnets, and recently transition ferromagnets has been demonstrated in several experiments.[20,29-34] A magnetic field induced by circularly polarized light due to the inverse optical Faraday effect, non-local transfer of angular momentum via spin currents, two-magnon scattering, demagnetization of different sites of the ferrimagnet (anti-ferromagnet), and exchange of the angular momentum between them due to different rates of magnetic moment change have been sought as the main contributors to all-optical switching (AOS) in such structures.[21,29,32,35-38]

Bulk and interface intrinsic and extrinsic spin-orbit interactions (SOI) have enabled the generation of pure spin currents leading to spin-transfer torques in different structures such as multi-layers of ferromagnets and paramagnets, thin Rashba ferromagnets, impure ferromagnets and semi-conductors, structures with non-zero Berry curvature, and etc.[39-54] Such spin currents and the torque exerted by them on the magnetization have been utilized in different configurations for switching or oscillation.[55-59] The intriguing observation of AOS in transition metal ferromagnets within specific structures containing paramagnets[30] has raised a debate for the role of SOI in dynamics under short-pulse optical excitations.

In this work, we discuss the effect of different spin-orbit interactions under the influence of short-pulse optical excitation to investigate the possibility of magnetization reversal of an out-of-plane magnetized structure. In section II, we qualitatively study the effect of different SOI mechanisms in the interested system. In section III, we present a numerical method to track the dynamics and distribution of the non-equilibrium diffusive transport of spin-polarized electrons. Compared to previous works that used a similar numerical method to explain the demagnetization, our approach includes exchange mechanisms and SOI. SOI causes the generation of spin currents with transverse (with respect to the magnetization) spin polarization, resulting in non-collinear spin distribution at each event (space-time point), which makes the inclusion of exchange interactions necessary. In section IV, we use the numerical method to demonstrate how the combination of transient currents, spin-Hall effect, and skew scattering can lead to magnetization reversal.

## II. SPIN-ORBIT INTERACTION MECHANISMS UNDER fs-LASER EXCITATION

The intrinsic role of SOI in demagnetization has been previously included in the models of Elliott-Yafet mechanisms considering the mixing of the spin states, and also in relativistic quantum thermodynamic for extracting the coherent response.[1,2,13,16,28] However, as the fs-laser excitation induces transient currents, the existence of different SOI mechanisms leads to transient spin currents and inversely generated charge currents.[5,7,8,14,17,18,22] Here, we are interested in structures with the out-of-plane (OOP) magnetization, and the shining laser to be propagating in the OOP direction. Such a structure is a major candidate for ultrafast high-density magnetic storage. Figure 1(a) shows the schematic of a ferromagnet (FM)/non-magnet (NM) bilayer, and the laser spatial absorption pattern. The absorption of fs-laser excites the electrons below the Fermi level. The conduction of such electrons occurs with different velocities than that of the electrons in the Fermi level and might be spin-dependent.[60-62] In addition, the inhomogeneity of the light absorption induces the inhomogeneous carrier density in the conduction band that leads to diffusive charge currents (if the material is FM, the currents are spin-dependent). Since the material is assumed to be magnetized in the *z* direction, the excited electrons initially carry spins parallel to the *z* direction.

In the absence of SOI mechanisms and magnetization chirality in the optically excited area and its neighboring region, the spin of the carriers remains parallel to the quantization axis, i.e. the *z* direction. It should be noted that the lack of transverse spin (spin polarization in the *xy* plane) means that even in the presence of high demagnetization and magnetic field induced by an inverse optical Faraday effect due to ellipticity of light polarization, there will be no deterministic magnetization reversal (random magnetic domains with reversed magnetization may form).

We define the spatially varying tensor of $\left[\Gamma^{C \to S}\right]_{s\sigma'c\sigma}(r,\varphi,z)$, which relates the charge currents to spin currents with certain spin polarizations ($(r,\varphi,z)$ indicate the cylindrical coordination). We also define $\left[\Gamma^{S \to C}\right]_{c\sigma s\sigma'}(r,\varphi,z)$ which indicates the charge currents induced by spin currents. The subscripts *c*, *s*, and $\sigma$ ($\sigma'$) refer to the directions of the charge current, spin current and spin polarizations, respectively. We qualitatively discuss below the effect of different

SOI mechanisms, and one would fill up the $\Gamma^{C \to S}$ and $\Gamma^{S \to C}$ tensors accordingly. The system symmetries (circular shape of the light spot and the perpendicular magnetization) imply that the net spin or charge accumulation in the $\hat{\varphi}$ direction will be zero. Therefore, the contribution of any SOI that leads to a spin current or charge current in the $\hat{\varphi}$ direction can be ignored, and the respective component in $\Gamma^{C \to S}$ or $\Gamma^{S \to C}$ can be set to zero.

The transient charge currents, parallel to the $z$ direction with a $z$-polarized spin current in the $z$ direction, do not induce any charge current ($\vec{S}(=S_z\hat{z}) \times \vec{J}^C(=J_z^C\hat{z}) = 0$) due to inverse SOI mechanisms, such as the inverse spin-Hall effect (ISHE) or inverse Rashba interaction at the interface of the FM/NM[63], where $\vec{S}$ is the spin polarization vector and $\vec{J}^C$ is the charge current vector. In addition, the Rashba interaction ($J_z^C\hat{z} \times \vec{n}(=\hat{z}) = 0$, where $\vec{n}$ is the normal to the interface) and the impurities (skew scattering in this case gives $S_z\hat{z} \times J_z^C\hat{z} = 0$) do not scatter this charge current, and do not give rise to any net spin current. The only possible effect of $J_z^C\hat{z}$ with the spin polarization of $S_z\hat{z}$ is an in-plane spin-current (in the radial direction) that is induced due to spin-Hall effect (SHE) in the bulk of the NM, where the polarization of this spin current will be in-plane and parallel to $\hat{\varphi}$, giving rise to an accumulation of in-plane spins (transverse spins) as shown in Fig. 1(b).

The in-plane transient charge currents flow in the radial direction $\vec{J}^C = J_r^C\hat{r}$, initially spin-polarized completely in the $z$ direction $\vec{S} = S_z\hat{z}$ (see Fig. 1(c)). If there are heavy metal impurities within the FM, the skew scattering will give rise to two different local spin currents. One of the spin currents will follow circular trajectories while being polarized in the $z$ direction. The net spin-accumulation due to this spin-current will be zero because of the circular trajectory (see Fig. 1(c)). The other spin-current flows parallel to the $z$ direction, while polarized in-plane and in the $\hat{\varphi}$ direction (see Fig. 1(c)). Due to this spin current, there will be a net transverse spin-accumulation towards the top surface and the interface between the FM and NM.

As discussed above, both the radial and OOP charge currents will acquire a transverse spin which are affected differently by SOI mechanisms. As can be seen in Fig. 1(d), the $J_z^S$ ($\vec{J}^S$ is the spin current) with $S_r$ or $S_\varphi$ will give rise to ISHE-induced charge currents parallel to $\hat{\varphi}$ and $\hat{r}$, respectively. $J_z^C$ with $S_r$ or $S_\varphi$ will induce a charge current parallel to $\hat{r}$ due to SHE in the NM. In addition, $J_r^C$ with $S_r$ or $S_\varphi$ will go through the skew scattering in the FM and give rise to two different spin currents. One spin current will have the same in-plane spin-polarization and flow in the $z$ direction. The other current will be polarized in the $z$ direction, while flowing in circular trajectories inducing zero net spin-accumulation.

The Rashba interaction in the interface induces direct or inverse scatterings of the in-plane charge and spin currents. As can be seen in Fig. 1(e), $J_r^C$ is scattered due to Rashba interaction and gives rise to a spin current parallel to $\hat{\varphi}$ with spin-polarization in the $\pm\hat{\varphi}$ direction. $J_r^S$ with $S_r$ induces a charge current parallel to $\hat{\varphi}$ due to the inverse Rashba interaction. Similarly, $J_\varphi^S$ with $S_\varphi$ induces a charge current parallel to $\hat{r}$. As mentioned above, the assumptions on the symmetry of the system indicate that the net spin and charge accumulations in the $\hat{\varphi}$ direction are zero. Therefore, the net contribution of Rashba interaction at the interface is zero. However, depending on the shape of the light spot (e.g. elliptical, rectangular, etc.) and the magnetization direction (e.g. unidirectional in-plane), the Rashba contribution can be non-zero.

In summary, transverse spin accumulation in a FM/NM bilayer with a perpendicularly magnetized FM under optical excitation is possible, and its spatiotemporal profile is in mutual correlation with the transient charge currents through the skew-scattering, interface Rashba, and bulk SHE mechanisms. In addition, the demagnetization is affected, but negligibly (see section IV), by the SOI mechanisms as the transient charge currents are modified by being scattered into spin currents with transverse spin polarization that go through different scattering types and rates.

## III. THEORETICAL MODEL

In order to demonstrate the possibility of optically-induced magnetization reversal, the study of the spin dynamics in the system under consideration is divided into two consecutive approaches. First, the non-equilibrium spatiotemporal spin density distribution is calculated through a transport model. The output of this approach is the spatiotemporal profile of moment $m_s$ (the spin polarization of the electrons below the Fermi energy $E_f$), as well as the density distribution of different spin components of the thermalized electrons at and above the Fermi level, $\vec{S}_{neq}$. Second, the dynamics of the magnetization $\vec{m}$ is calculated by using the Landau–Lifshitz–Gilbert (LLG) equation. The torque exerted by $\vec{S}_{neq}$ on $\vec{m}$, denoted by $T_m$, is included in the LLG equation, and is nonzero if $\vec{m}$ and $\vec{S}_{neq}$ are non-collinear.

In order to determine the spatiotemporal distribution of $\vec{S}_{neq}$ and $m_s$, we model the transient spin-dependent transport of the electrons thermalized by the optical excitation in the presence of the SOI mechanisms. We define the carrier source terms including the ones that are optically excited ($\Lambda_{opt}$), and the scattered part of the transported carriers ($\Lambda_{sc}$) (all the scatterings excluding SOI; SOI scatterings will be added separately in the diffusion equation) for all the possible energies and spin transport channels as,

$$\Lambda_\alpha(E,r,\varphi,z,t) = \Lambda_{opt,\alpha}(E,r,\varphi,z,t) + \Lambda_{sc,\alpha}(E,r,\varphi,z,t), \quad (1)$$

where $\alpha$ is the spin channel which can be $r$, $\varphi$, $z_\uparrow$, and $z_\downarrow$ (the spin-channel parallel to the quantization axis $z$ is divided into up and down channels for convenience), whereas $E$ is the indicator of the channel energy. In the rest of the formulations, $(E,r,\varphi,z,t)$ is dropped for simplicity, and if needed only the important parameters are mentioned.

In Eq. (1), $\Lambda_{opt,\alpha} = \Lambda_{qax,\alpha} + \Lambda_{acc,\alpha} \delta(E - (E_f + E_{ph}))$, where $\Lambda_{qax,\alpha} = \Lambda_{las}(t) \Upsilon(\alpha, r, z)(\delta_{\alpha, z_\uparrow} + \delta_{\alpha, z_\downarrow})$ indicates the electrons optically excited from the states below the Fermi level with a spin polarization parallel to the quantization axis. $\Lambda_{acc,\alpha}(E_f + E_{ph}) = \Lambda_{acc,\alpha}(E_f) \Lambda_{las}(t) \Upsilon(\alpha, r, z)$ indicates the electrons optically excited from the Fermi level (therefore the excited electrons in this term are at $E_f + E_{ph}$) which can have a non-collinear spin polarization with respect to the quantization axis due to the existence of SOI mechanism. $\Lambda_{las}(t)$ is only a function of time which represents the power of the laser and is assumed to be Gaussian, whereas $\Upsilon(\alpha, r, z)$ represents the spatial functionality of the laser absorption. $\Upsilon(\alpha = z_\uparrow, z) = n_{d,\uparrow} e^{-z/\lambda} f(r)$, $\Upsilon(\alpha = z_\downarrow, z) = n_{d,\downarrow} e^{-z/\lambda} f(r)$, and $\Upsilon(\alpha = x(y), r, z) = f(r) e^{-z/\lambda}$, where $n_{d,\uparrow(\downarrow)}$ is the percentage of the up (down) spin excited above the Fermi level ($E_f$),[64] $f(r)$ is the radial laser absorption function, and $\lambda$ is the absorption length. $\Lambda_{sc,\alpha}(E,r,\varphi,z,t) = \sum_{\alpha'} \sum_{E'} P(\alpha, \alpha', E, E', r, \varphi, z, t) n_{\alpha'}(E', r, \varphi, z, t) \Delta E$.

$P(\alpha,\alpha',E,E',r,\varphi,z,t)$ indicates the scattering rate from a state with an energy $E'$, and spin channel $\alpha'$ into a state with an energy $E$ and spin channel $\alpha$. $n_\alpha(E,r,\varphi,z,t)$ and $\tau_{tot}(E,r,\varphi,z,t)$ are the carrier density and the total scattering rate of the spin channel $\alpha$ and energy $E$, respectively, at a specific time and space. It is worth noting that $1/\tau_{tot}(E',r,\varphi,z,t) = \sum_\alpha \sum_E P(\alpha,\alpha',E,E',r,\varphi,z,t)$.

The flux of carriers as a function of time and space, can be written as

$$\vec{\Phi}_\alpha(E,r,\varphi,z,t) = \int_0^{+\infty} dr' \int_0^{2\pi} d\varphi' \int_{-\infty}^{+\infty} dz' \int_{-\infty}^{t} \Lambda_\alpha(E,r',\varphi',z',t') \vec{\Xi}_\alpha(E,r,\varphi,z,t/E,r',\varphi',z',t') dt', \quad (2)$$

where $\vec{\Xi}_\alpha$ is the propagator based on Boltzmann formalism, and its derivation includes the spin and energy dependent elastic and inelastic scatterings, as well as velocities.[14,18] If we ignore the dependence on $\varphi$ which is allowed due to the symmetries, we have modified $\vec{\Xi}_\alpha$ derived in the literature for the case of infinite plane, to account for finite and meshed radial dimension (refer to Appendix for details). $\vec{\Xi}_\alpha$ depends on the total scattering rate $\tau_{tot}$ which comprises of several types of scattering, and can be written as

$$\frac{1}{\tau_{tot}} = \frac{1}{\tau_{d,in}} + \frac{1}{\tau_{d,el}} + \frac{1}{\tau_{sf}} + \frac{1}{\tau_{sd}} + \frac{1}{\tau_{ss}}, \quad (3)$$

where $\tau_{d,in}$ is the inelastic, $\tau_{d,el}$ is the elastic, and $\tau_{sf}$ is the spin-flip scattering. $\tau_{sd}$ indicates the exchange interaction between the bounded spin carriers (below $E_f$) and the non-equilibrium conduction carriers (at and above $E_f$), whereas $\tau_{ss}$ stands for the exchange interaction among the non-equilibrium conduction carriers. The relation of $P$ to different scattering terms ($\tau_{d,in}$, $\tau_{d,el}$, $\tau_{sf}$, $\tau_{sd}$, and $\tau_{ss}$) are shown in the Appendix.

The carrier density in each energy and spin channel is derived by solving the diffusion equation

$$\frac{\partial n_\alpha(E)}{\partial t} + \frac{n_\alpha(E)}{\tau_{tot}(E)} = \left(-\nabla \cdot \vec{\Phi}_\alpha(E) + \Lambda_\alpha(E) + \vartheta_\alpha(E) + \vartheta'_\alpha(E)\right), \quad (4)$$

where $\vartheta_\alpha(E)$ indicates the contribution of the SOI mechanisms, and can be defined based on the SOI induced currents discussed in section II, as $\vartheta_\alpha(E) = (\nabla \cdot \vec{J}^S) S_\alpha^{SC}$, where $S_\alpha^{SC}$ is the spin polarization of the scattered spin/charge current. We can write $\vec{J}^C = \vec{\Phi}_\alpha$ and use the tensors $\Gamma^{C \to S}$ and $\Gamma^{S \to C}$ to calculate $\vec{J}^S$ and $\vec{S}^{SC}$. To conserve the number of conducting carriers, we have included $\vartheta'_\alpha(E) = -(\nabla \cdot \vec{J}^S) \times \left\{\sqrt{\sum_{\alpha'} (S_{\alpha'}^{SC})^2 (1-\delta_{\alpha,\alpha'})}\right\} \times n_\alpha(E)/|n_\alpha(E)|$ in Eq. (4).

Knowing the spin density of the thermalized electrons above $E_f$ ($n_\alpha$) and below $E_f$ ($n_{<E_f, z_{\uparrow(\downarrow)}}$) allows to determine the profile of $m_s$ and $\vec{S}_{neq}$. Solving Eq. (4) provides $n_\alpha$, and $n_{<E_f, z_{\uparrow(\downarrow)}}$ can be determined by

$$\frac{\partial n_{<E_f, z_\uparrow(z_\downarrow)}}{\partial t} = \left(-\int \Lambda_{qax, z_\uparrow(z_\downarrow)}(E) dE + \int \left[1 - \sum_{\alpha'} \int P(\alpha', z_\uparrow(z_\downarrow), E', E) dE'\right] n_{z_\uparrow(z_\downarrow)}(E) dE\right).^{14,18}$$

Therefore, we can write

$$m_s = 2\mu_B \left(n_{<E_f, z_\uparrow} - n_{<E_f, z_\downarrow}\right) + M_s, \quad \vec{S}_{neq} = 2\mu_B \int \left[n_x \hat{x} + n_y \hat{y} + (n_{z_\uparrow} - n_{z_\downarrow})\hat{z}\right] dE, \quad (5)$$

where $\mu_B$ is the Bohr magneton, and $M_s$ is the saturation magnetization of FM at room temperature. Since the torque on the non-equilibrium conduction spins are already included through the exchange terms in the scattering ($\tau_{sd(ss)}$), we calculate the torque on the bounded magnetization $\vec{m} = m_s \hat{z}$ separately. The torque on $\vec{m}$ can be written as $\vec{T}_m = -\frac{1}{\Delta_{sd}} \vec{m} \times \vec{S}_{neq}$. We can define the dynamic of $\vec{m}$ using the LLG equation $\frac{d\vec{m}}{dt} = \gamma \vec{m} \times \vec{H} + \vec{T}_m + \alpha \vec{m} \times \frac{d\vec{m}}{dt}$, where $\gamma$ is the gyromagnetic ratio, $\alpha$ is the Gilbert damping constant, and $\vec{H} = \vec{H}_d + \vec{H}_{ex} + \vec{H}_{IF} + \vec{H}_{anis} + \vec{H}_{ext}$. $\vec{H}_d$ is the dipolar field, $\vec{H}_{ex}$ is the exchange field due to the possible non-collinear spatial functionality of $\vec{m}$ due to both isotropic and anisotropic Dzyaloshinskii-Moryia interaction (DMI)[65,66], $\vec{H}_{IF}$ is the field induced by the inverse optical Faraday effect due to ellipticity of optical excitation, $\vec{H}_{anis}$ is the anisotropy field, and $\vec{H}_{ext}$ is the external field.

## IV. NUMERICAL RESULTS AND DISCUSSION

We utilize the SOI mechanisms in section II and the theoretical model of section III in order to demonstrate the possibility of AOS in a magnetic system, which does not have any ferrimagnetic order and its FM layer has OOP crystalline anisotropy ($\vec{H}_{anis} = \frac{2K_u}{\mu_0 m_s}(\vec{m} \cdot \hat{z})\hat{z}$, where $K_u$ is the anisotropy constant and $\mu_0$ is the vacuum permeability). With assuming a circular symmetry of the system (circularly isotropic impurity distribution within the FM and circular shape of the light boundary) and zero DMI ($\vec{H}_{ex} = A\nabla^2 \vec{m}$, where $A$ is the isotropic exchange constant), we need to consider the spin dynamics in the spatial variation only in the radial and parallel to the $z$ directions. In most of the experimental situations, the light spot radius ($R_{las}$) is much bigger than the thickness of the stack ($d_{tot}$). For reliable calculations we need the cell sizes in the $z$ and $r$ directions to be comparable ($N_z$ ($N_r$) is the number of mesh cells in the $z$ (radial) direction and $D_m = d_{tot}/N_z$ ($R_m = R_{las}/N_r$) is the respective mesh size). The experimental situation ($R_{las} \gg d_{tot}$) while keeping $D_m \sim R_m$ requires matrix sizes and calculation times that are beyond the limit of our numerical method. Therefore, we focus on $R_{las}$ in the same order as $d_{tot}$. In addition, such a dimension is more relevant in the AOS application of high density magnetic storage. To simplify the calculations further, we approximate the $\tau_{tot}$ as a constant ($\frac{1}{\tau_{tot}} = \frac{1}{\tau_{d,in}} + \frac{1}{\tau_{d,el}} + \frac{1}{\tau_{sf}}$) while calculating $\vec{\Xi}_\alpha$.

For the inelastic scattering $\tau_{d,in}$ and carrier velocity $v_\alpha(E,z)$, we use the values of Fe for the FM layer, and use values of Pt for the NM layer.[61,62] We assume $\tau_{d,el} = 4/3\tau_{d,in}$, $\tau_{sf}$ to be 2000 fs and 20 fs for the FM and the NM (in the order of reported magnitudes for ferromagnets such as Co and Fe, and non-magnets such as Pt and W),[67] respectively. $\Delta_{sd}$ ($\Delta_{ss}$) are assumed to be 1 fs (1 fs) for the FM and $\infty$ fs (1 fs) for the NM. $\Delta_{sd} = \hbar/\delta_{sd}$, where $\delta_{sd}$ is the exchange splitting and the reported value for Fe is around 2 eV,[68] making 1 fs a reasonable choice of $\Delta_{sd}$ for the calculations not to overestimate the results. The values for $\Delta_{ss}$ in FM and NM is chosen to account for the exchange interaction between the conduction electrons (based on the mean-field Weiss model), and at the same time not to overestimate the results for the transverse spin. The thicknesses of the FM and NM layers are assumed to be $t_{FM}$ = 1.5 nm and $t_{NM}$ = 2.5 nm, respectively. The other parameters are assumed to be $f(r) = exp(-2r^2/(aR_{las})^2)(1-\Theta(r-R_{las}))$ ($\Theta$ is a Heaviside function and a coefficient, $a$ is assumed to be 2), $\Lambda_{las}(t) = P_{las} exp(-2(t-\varsigma/2)^2/\varsigma^2)\Theta(t)$ mW/m$^2$ ($\varsigma$ is the full width at half maximum of the Gaussian-shape laser pulse, and $P_{las}$ is its peak of effectively absorbed power),

$\varsigma = 50$ fs, $E_{las} = \int_{-\infty}^{+\infty} P_{las} dt = 1$ mJ/m$^2$, $\Gamma_{sh} = -0.3$ is the spin-Hall angle of the NM (the values reported for Ta and W can be $-0.3$, the reported values for Pt are $\sim 0.1$)[58,69,70], $\Gamma_{sk} = -0.1$ is the skew scattering angle in the FM (a value as high as $-0.24$ is reported by skew scattering of Bi impurities in Cu which has negligible intrinsic spin-Hall angle)[71], $\lambda = 15$ nm, $n_{d,\uparrow} = 0.5$, $n_{d,\downarrow} = 0.5$, $\vec{H}_{ext} = 0$, $K_u = 10^6$ J/m$^3$ (based on values reported for $L_{10}$ FePt as well as multilayers such as Ta/CoFeB/MgO), $\gamma = 2.2 \times 10^5$ m/(A·s), $\alpha = 0.1$, $R_{las} = 5$ nm, $N_z = 40$, and $N_r = 10$. In the calculations we use the above parameters values unless otherwise specified.

The direction of the relativistic part of the inverse optical Faraday field is determined by the helicity of the optical excitation, while it has a non-trivial dependence on the helicity for the non-relativistic part.[72-74] However, based on experiments in systems similar to the ones studied here, we can assume $\vec{B}_{IF} = -1\hat{z}$ T for a possible assistance of magnetization reversal (minus sign corresponds to a left-handed circularly polarized laser exciting a FM (Co)/NM (Ta) bilayer).[75] As this amplitude for $\vec{B}_{IF}$ contributes to the magnetization dynamics in a larger time scale ($\sim$ 1 ps) than the one due to spin-transfer torque ($\sim$ 10 fs) induced by the transverse spin accumulation, it can be set to zero. This means that the reversal process presented here does not depend on the polarization of the shining light, however, it can be enhanced if $\vec{B}_{IF}$ is large enough and in the assisting direction (if switching from $+z$ to $-z$, the sign of $\vec{B}_{IF}$ should be negative).

We solve Eq. (4) self-consistently at $\varphi = \pi/2$ (corresponding to $\hat{r} = \hat{x}$ and $\hat{\varphi} = \hat{y}$). Figure 2(a) shows the spatial average (over $0 < r < R_{las}$ and $t_{NM} < z \leq d_{tot}$) of the temporal evolution of $\vec{S}_{neq}$ and the change of $m_s$ in the FM layer ($\vec{S}_{avg}$ and $\Delta m_z$, respectively) for up to 160 fs. The demagnetization (reduction of $m_s$) and the presence of transverse spins ($\vec{S}_{neq} \cdot (\hat{r} + \hat{\varphi}) \neq 0$) are evident from Fig. 2(a). Figures 2(b-d) show the radial average (over $0 < r < R_{las}$) of the spatial distribution of $\vec{S}_{neq}$ in the $\hat{r}$, $\hat{\varphi}$, and $\hat{z}$ directions. It should be noted that the irregular oscillations in the z direction of some spatiotemporal regions in Figs. 2(b-e) are due to finite space and time mesh dimensions. However, we have checked that for the system under consideration, the mesh sizes used here give reliable results and main behaviors of the signals are captured properly (the artifact oscillations are averaged to negligible values). The non-equilibrium spin texture formed in the NM is originated mainly in the radial *rz*-variant SHE-induced spin currents which also diffuse along the *z* and *r* directions (going through reflections from the bottom boundary and diffusion into the FM). The spin texture in the FM layer is further affected by the OOP *rz*-variant skew-induced spin currents which also diffuse along the *z* and *r* directions (going through reflections from the top boundary and diffusion into the NM). In addition, the exchange interactions ($\tau_{sd}$ and $\tau_{ss}$) are responsible for the correlation between the three coordinates of the non-equilibrium spins.

The dip of $S_z$ in Fig. 2(a) at ~ 50 fs is the direct effect of the optical excitation in FM and diffusion of electrons into the NM which are initially spin-polarized along the magnetization (see also Fig. 2(d)). The velocities and scattering rates are different for spin-up and spin-down, resulting in different diffusion rates, which leads to a non-zero temporal profile of $S_z$ in FM. The *sd* exchange interaction which turns the transverse-spins along the magnetization (*z* direction), affects $S_z$, however the effect is negligible due to $|\vec{S}_{neq} \cdot (\hat{r} + \hat{\varphi})| \ll S_z$ (see also Fig. 3(b)). The decay in all the components of non-equilibrium spins is expected as the transient charge and spin currents equilibrate. The demagnetization ($\Delta m_z$) becomes saturated as long as another temperature bath (e.g. lattice) is not included (see Fig. 2(a)). We focus on the transient currents in the presented time scale, where other temperature bathes can be neglected. Figure 2(e) shows the radial average of the spatial distribution of $\Delta m_z$. The demagnetization process is similar to the results from the models excluding the SOI mechanisms, where the spin-up and spin-down optically excited electrons diffuse with different velocities and scattering rates into the adjacent NM, causing the demagnetization.[8,14] Here, we have non-zero $\vec{S}_{neq} \cdot (\hat{r} + \hat{\varphi})$, but $\left| \vec{S}_{neq} \cdot (\hat{r} + \hat{\varphi}) \right| \ll S_z$, therefore we expect similar demagnetization process.[8,14]

In Figs. 3(a-e) we present the effect of different parameters of the system on the temporal evolution of $\vec{S}_{neq}$ and $\Delta m_z$. Figure 3(a) shows the dynamics of $\vec{S}_{neq}$ for three different thicknesses of the NM (2.5 nm (case 1), 2.0 nm (case 2), and 3.0 nm (case 3)). With increasing $t_{NM}$ the demagnetization increases (till it saturates) as the diffused electrons from FM into NM flow back less into FM. In addition, by increasing $t_{NM}$ the amount of spin accumulation due to SHE increases, and consequently its diffusion into the FM leads to an increase in the amplitude of $S_\varphi$. Figure 3(b) shows the spin dynamics for three different strengths of the exchange interactions ($\Delta_{sd}$ ($\Delta_{ss}$) = 1 (similar to case 1), 2, and 5 fs). Higher values of $\Delta_{sd}$ ($\Delta_{ss}$) mean less scattering of $\left( \vec{S}_{neq} \cdot (\hat{r} + \hat{\varphi}) \right)$ into $S_z$ or $m_z$, resulting in higher amplitudes of $S_\varphi$. However, as $\left| \vec{S}_{neq} \cdot (\hat{r} + \hat{\varphi}) \right| \ll S_z$, change in demagnetization and $S_z$ should be negligible as can be also seen in Fig. 3(b). Figure 3(c) shows the spin dynamics for four different cases of skew scattering ($\Gamma_{sk}$ = -0.3 (case 4), -0.1 (similar to case 1), 0 (case 5), and 0.3 (case 6)). The effect of skew scattering dominates the initial stages of the spin dynamics where the SHE induced spins have not diffused into the FM yet. Therefore, the effect of the changes in the value and sign of $\Gamma_{sk}$ is pronounced in the initial stages of the $\vec{S}_{neq} \cdot (\hat{r} + \hat{\varphi})$ (see $S_\varphi$ in Fig. 3(c)). Again, the effect of $\Gamma_{sk}$ value and sign on $\Delta m_z$ and $S_z$ is negligible as $\left|\vec{S}_{neq} \cdot (\hat{r} + \hat{\varphi})\right| \ll S_z$.

Figure 3(d) shows the spin dynamics for three different cases of the $f(r)$ lineshape coefficient $a$ ($a = 1$ (case 7), 2 (similar to case 1), 4 (case 8)). The smaller $a$ indicates higher concentration of light in the center, leading to an increase of spin accumulation due to the skew scattering and the SHE in NM at earlier times. The larger $a$ is, the higher light absorption becomes and the higher amplitude signals have. It should be noted that the amplitude of the signals can be tuned by changing the laser power, while the lineshapes and the relative values of signal components remain unchanged (as long as a conduction channel is not saturated at higher laser powers). Figure 3(e) shows the spin dynamics for three different values of absorption length ($\lambda = 5$ (case 9), 15 (similar to case 1), and 25 nm (case 10)). For smaller $\lambda$, the absorption of light is steeper in the $z$ direction leading to higher diffusive currents, which consequently results in higher diffused SHE-induced spin-accumulation into FM at later times. It can also be noticed that for smaller values of $\lambda$, the amplitude of $S_z$ increases while the demagnetization decreases.

For the cases 1 to 10 presented in Figs. 3(a) and 3(c-e), we explore the possibility of magnetization reversal by solving the LLG dynamics of the magnetization $\vec{m}_{avg}$, while assuming that $\vec{S}_{neq}$ (see Eq. (5)) keeps its coordination throughout the time. To be consistent, we assume a laser energy that causes $min(m_s) = M_s + min(\Delta m_z) = 0.1$ μ$_B$/atom (the saturation magnetization of FM, $M_s$ is assumed to be 0.65 μ$_B$/atom) for each of the cases 1 to 10, and then we vary $\Delta_{sd}$ to find its upper threshold for achieving magnetization reversal (see Fig. 3(f)). The existence of non-equilibrium transvers spin ($S_r$ and $S_\varphi$) without longitudinal one ($S_z$) brings the magnetization towards in-plane but does not switch it. Therefore, both the amplitude ratio of $S_\varphi$ to $S_z$ (it should be noted that the amplitudes can be scaled by laser power, therefore only ratios and relative profiles are important), and overlap details of their temporal lineshapes are important to determine the reversal possibility for a particular $\Delta_{sd}$. The possibility of reversal for all the considered cases at a reasonable value of $\Delta_{sd}$, and few other important points can be inferred from Fig. 3(f) as follows. The presence, sign, and amplitude of skew scattering (with respect to the spin-Hall angle) can modulate the reversal possibility considerably (cases 1, 4, 5, and 6, see also Fig. 3(c)), due to the modulations it causes in the initial temporal stages of $S_\varphi$ and the overlap of $S_\varphi$ lineshape with $S_z$, which in turn determine the rate of reversal. From cases 1, 7, and 8, it can be inferred that the smaller the *f(r)* lineshape coefficient *a* is (a lower value of *a* means a higher concentration of light in the center), the easier the magnetization reversal becomes. In the overlap region of $S_\varphi$ and $S_z$ lineshapes, the relative amplitude of $S_\varphi$ with respect to $S_z$ is higher for the case 7 (*a* = 1) with respect to cases 1 (*a* = 2) and 8 (*a* = 4) (see Fig. 3(d)). Similar discussion is applicable for the effect of $\lambda$, where the cases 9 and 10 in Fig. 3(f) indicate an easier reversal for a larger $\lambda$. For a smaller $\lambda$, the absorption of light will be steeper in the *z* direction, resulting in higher spin-accumulation due to SHE in later times (where $S_z$ is suppressed), which makes the magnetization remain in-plane up to smaller values of $\Delta_{sd}$. In conclusion, by manipulating the SOI strengths and transient currents (determined by material structures as well as laser power and its spatiotemporal pattern), the non-equilibrium signal shape can be modified in order to manipulate the magnetization reversal for AOS applications.

As Figs. 3(d) and 3(f) suggest, the *f(r)* lineshape coefficient *a* modifies the signals ($\vec{S}_{neq}$ and $\Delta m_z$) and the reversal possibility (compare the cases 1 and 7) more significantly than the other parameters. Figure 4(a) shows that the amplitude of $S_\varphi$ is enhanced at the earlier stage ($0 < t < 40$ fs) by decreasing *a*, and $|S_\varphi/S_z|$ increases. Consequently, the reversal possibility in Fig. 4(b) generally increases with decreasing *a*. Figures 4(c,d) and 4(e,f) show the spatiotemporal distribution of $\vec{S}_{neq}$ in the $\hat{\varphi}$ direction in the NM ($a=1$ and 2) and FM ($a=1$ and 2), respectively. For smaller *a*, the positive $S_\varphi$ spin accumulation in NM shifts towards the center of the laser spot (see Fig. 4(c,d)), which results in more diffusion of positive $S_\varphi$ from NM into FM at the earlier stage (see Fig. 4(e,f)), leading the profile of $S_\varphi$ to be more positive in FM.

Although the above calculations are based on several approximations, the results indicate the possibility of magnetization reversal through the coexistence of optically excited transport and different SOI mechanisms. In order to achieve a more accurate model, the Elliott-Yafet mechanisms should be included, the dependence of $\vec{\Xi}_\alpha$ in time and all the three dimensions should be considered, the coefficients regarding the SOI mechanisms as well as $\Delta_{sd}$, and $\Delta_{ss}$ should be derived through first-principle calculations, and the magnetization dynamics should be solved micromagnetically. In addition, Monte Carlo methods can be implemented to perform calculations for larger systems.

## V. CONCLUSIONS

We have discussed the effect of SOI mechanisms under the fs optical excitation. In perpendicularly magnetized ferromagnetic thin films adjacent to a non-magnetic heavy metal layer, the skew scattering and spin-Hall effect give rise to the generation of transverse spins, which in turn exerts torques on the magnetization. Such torques can cause deterministic magnetization reversal, enabling all-optical switching in ferromagnetic transition metals. In order to optimize the structure for such a reversal mechanism, the impurity type, the bulk of the heavy metal, and the Rashba coefficient at the interface should be designed so as to have their effects superimpose constructively. In addition, the shape of the light spot and the light absorption coefficients, which determine the transient charge currents, should be optimized in order to facilitate the magnetization reversal process.

## ACKNOWLEDGEMENTS

This work was supported by the National Research Foundation (NRF), Prime Minister's Office, Singapore, under its Competitive Research Programme (CRP Award No. NRF CRP12-2013-01). H.Y. is a member of the Singapore Spintronics Consortium (SG-SPIN).

## APPENDIX

The propagator used to derive the flux of carriers $\vec{\Phi}_\alpha(E,r,\varphi,z,t)$ in Eq. (2) can be written for the three dimensional case as

$$\vec{\Xi}_\alpha = \frac{Z}{2(t-t')^2} exp\left[-(t-t')(\mathrm{T})/(Z)\right]\Theta(t-t'-|Z|)\Theta(|Z|/(t-t')-\theta^-)\Theta(\theta^+ - |Z|/(t-t'))$$
$$\left[cos((\theta^+ + \theta^-)/2)\hat{z} + sin((\theta^+ + \theta^-)/2)\hat{r}\right],$$

where $Z = \int_{z'}^{z} \frac{dz''}{v(z'')}$, $T = \int_{z'}^{z} \frac{dz''}{\tau_{tot}(z'')v(z'')}$, $\theta^{+} = arctan\left(\left|\frac{r-r'-\Delta r/2}{z-z'}\right|\right)$ and

$\theta^{-} = arctan\left(\left|\frac{r-r'+\Delta r/2}{z-z'}\right|\right)$ ($\Delta r$ is the radial mesh size).

At each point of time and space, the scattering rate from a state with an energy $E'$, and spin channel $\alpha'$ into a state with an energy $E$ and spin channel $\alpha$, is determined by $P(\alpha,\alpha',E,E',r,\varphi,z,t)$ which is related to different scattering mechanisms. The parts of $P$ that are related to $\tau_{sd}$ and $\tau_{ss}$ can be written as $P_{sd}(\alpha,\alpha',E,E') = \frac{1}{\Delta_{sd}}\frac{m_{\beta}^2}{m_T^2}\varepsilon_{\alpha\beta\alpha'}\delta(E-E')$ and

$P_{ss}(\alpha,\alpha',E,E') = \frac{1}{\Delta_{ss}}\frac{n_{\beta}^2}{n_T^2}\varepsilon_{\alpha\beta\alpha'}\delta(E-E')$. $\Delta_{sd(ss)}$ are the scattering rates of the two aforementioned exchange interactions. $m_T^2 = \sum_{\alpha} m_{\alpha}^2$ and $n_T^2 = \sum_{\alpha} n_{\alpha}^2$, where $\alpha$ sums over the spin channels, and $m_{\alpha}$ is the $\alpha$-component of the magnetization. Based on $P_{sd(ss)}$, we can express $\tau_{sd(ss)}$ as $\frac{1}{\tau_{sd(ss)}} = \sum_{E'}\sum_{\alpha'} P_{sd(ss)}(\alpha',\alpha,E',E)$. $\Delta_{sd}$ depends on the magnetization of the carriers below $E_f$, $\frac{1}{\Delta_{sd}} \propto (n_{z\uparrow} - n_{z\downarrow})_{E<E_f}$, and $\Delta_{ss}$ depends on the details of the density distribution of the different spin components of the conducting electrons. However, in our calculations we approximate both of them to be constant values.

The part of $P$ regarding the elastic and inelastic scattering is

$P_{d,in} + P_{d,el} = \left[\delta_{\alpha,\alpha'}\frac{\Theta_{[E'-\Delta E_{max},E']}(E)}{\Delta E_{max}} + \chi\frac{\Theta_{[0,\Delta E_{max}]}(E)}{\Delta E_{max}}\right]\frac{1}{\tau_{d,in}}(1-\delta(E-E')) + \frac{1}{\tau_{d,el}}\delta_{\alpha,\alpha'}\delta(E-E')$

,[14,18] where $\Delta E_{max}$ is the maximum of the energy scattering for the inelastic scattering, and $\Theta_{[...]}$

is unity in the range indicated in brackets and zero otherwise. $\chi$ is defined to account for the excitation from $E_f$, which is a mixture of equilibrium electrons spin-polarized parallel to the quantization axis and non-equilibrium electrons polarized according to $n_\alpha$. $\chi$ can be written as

$$\chi = \left( Q \cdot C \cdot \frac{n_\alpha^2}{n_T^2} + Q \cdot (1-C) \cdot \frac{m_\alpha^2}{m_T^2} \right), \quad \text{where} \quad Q = 0.5\left(\delta_{\alpha,z_\uparrow} + \delta_{\alpha,z_\downarrow}\right) + \left(1 - \delta_{\alpha,z_\uparrow} - \delta_{\alpha,z_\downarrow}\right) \quad \text{and}$$

$0 \leq C \leq 1$. The parameter $C$ accounts for the ratio of the electrons polarized transverse and parallel to the $z$ axis. The value of $C$ is a dynamic value with respect to time and depends on the density of states and the details of non-equilibrium spin of the thermalized electrons at the Fermi level. In the calculations, we use a small value of $C = 0.01$ not to overestimate the results for $S_{r,\varphi}$. Finally, the part of $P$ regarding the spin-flip scattering can be written as

$$P_{sf} = \frac{1}{\tau_{sf}} \delta_{\alpha,\alpha'} \left(1 - \delta_{\alpha',z_\uparrow} - \delta_{\alpha',z_\downarrow}\right) + \frac{1}{\tau_{sf}} \left(1 - \delta_{\alpha,\alpha'}\right) \cdot \left(\delta_{\alpha',z_\uparrow} + \delta_{\alpha',z_\downarrow}\right).$$

**Figure captions**

**Fig. 1.** (a) The schematic of a FM/ NM bilayer and the laser spatial absorption pattern. The points in the FM are the impurities. (b) Top and side view of the in-plane spin-current (in the radial direction) with spin polarization along $\hat{\varphi}$, induced due to SHE in NM from $J_{c,z}\hat{z}$ with spin polarization of $S_z\hat{z}$. (c) Spin currents due to skew scattering of $J_{c,r}\hat{r}$ with spin polarization of $S_z\hat{z}$. Panels marked by numbers 1 and 2 refer to the two possible spin currents. The most right panel shows the side view of the spin current shown in panel 2. (d) ISHE-induced charge currents due to $J_{c,z}$ with transverse spin polarizations of $S_r$ or $S_\varphi$. (e) Spin current parallel to $\hat{\varphi}$ induced by $J_{c,r}$ due to the Rashba interaction at the FM/NM interface. The green (dashed black) arrows indicate direct (inverse) scattering. Black, red, blue, and pink arrows indicate charge current, spin polarization before scattering, spin current, and spin polarization after scattering, respectively. Dark blue (dark green) balls are non-scattered (scattered) electrons.

**Fig. 2.** (a) Spatial average (average over $0 < r < R_{las}$ and $t_{NM} \leq z < d_{tot}$) of the temporal evolution of $\vec{S}_{neq}$ and the change in $m_s$ in the FM layer ($\vec{S}_{avg}$ and $\Delta m_z$, respectively) up to 160 fs. (b-d) The radial average (average over $0 < r < R_{las}$) of the spatiotemporal distribution of $\vec{S}_{neq}$ in the $\hat{r}$, $\hat{\varphi}$ and $\hat{z}$ directions, respectively. (e) The radial average of the spatiotemporal distribution of the change in $m_s$, $\Delta m_z$.

**Fig. 3.** (a) $S_\varphi$, $S_z$ and $\Delta m_z$ for $t_{NM}$ = 2.5 nm (continuous line, case 1), 2 nm (dashed line, case 2), and 3 nm (dotted line, case 3). (b) $S_\varphi$, $S_z$ and $\Delta m_z$ for $\Delta_{sd}$ ($\Delta_{ss}$) = 1 fs (continuous line), 2 fs

(dashed line), and 5 fs (dotted line). (c) $S_\varphi$, $S_z$ and $\Delta m_z$ for $\Gamma_{sk} = -0.3$ (dashed line, case 4), $\Gamma_{sk} = -0.1$ (continuous line), $\Gamma_{sk} = 0$ (dash-dotted line, case 5), and $\Gamma_{sk} = 0.3$ (dotted line, case 6). (d) $S_\varphi$, $S_z$ and $\Delta m_z$ for $a = 1$ (dashed line, case 7), $a = 2$ (continuous line), and $a = 4$ (dotted line, case 8). (e) $S_\varphi$, $S_z$ and $\Delta m_z$ for $\lambda = 5$ nm (dashed line, case 9), $\lambda = 15$ nm (continuous line), and $\lambda = 25$ nm (dotted line, case 10). For (a-e), $t_{FM} = 1$ nm, $N_r = 5$, and the continuous line represents a similar case ($t_{NM} = 2.5$ nm, $\Delta_{sd}$ ($\Delta_{ss}$) = 1 fs, $\Gamma_{sk} = -0.1$, $a = 2$, and $\lambda = 15$ nm). (f) The value of $\Delta_{sd}$ that the magnetization reversal becomes possible while we keep $M_s + min(\Delta m_z) = 0.1$ $\mu_B$/atom, for the cases 1 to 10 (the error bar for $\Delta_{sd}$ is ±0.01 fs and dot sizes reflect this error bar).

**Fig. 4.** (a) $S_\varphi$ and $S_z$ for $a = 1$, 1.2, 1.5, and 2. (b) The value of $\Delta_{sd}$ that the magnetization reversal becomes possible while we keep $M_s + min(\Delta m_z) = 0.1$ $\mu_B$/atom, for the cases in (a) (the error bar for $\Delta_{sd}$ is ±0.01 fs and dot sizes reflect the error bar). (c-d) The spatiotemporal distribution of $\vec{S}_{neq}$ in the $\hat{\varphi}$ direction in the NM (averaged over $0 < z \leq d_{NM}$) for $a = 1$ and $a = 2$, respectively. (e-f) The spatiotemporal distribution of $\vec{S}_{neq}$ in the $\hat{\varphi}$ direction in the FM (average over $d_{NM} < z < d_{tot}$) for $a = 1$ and $a = 2$, respectively.

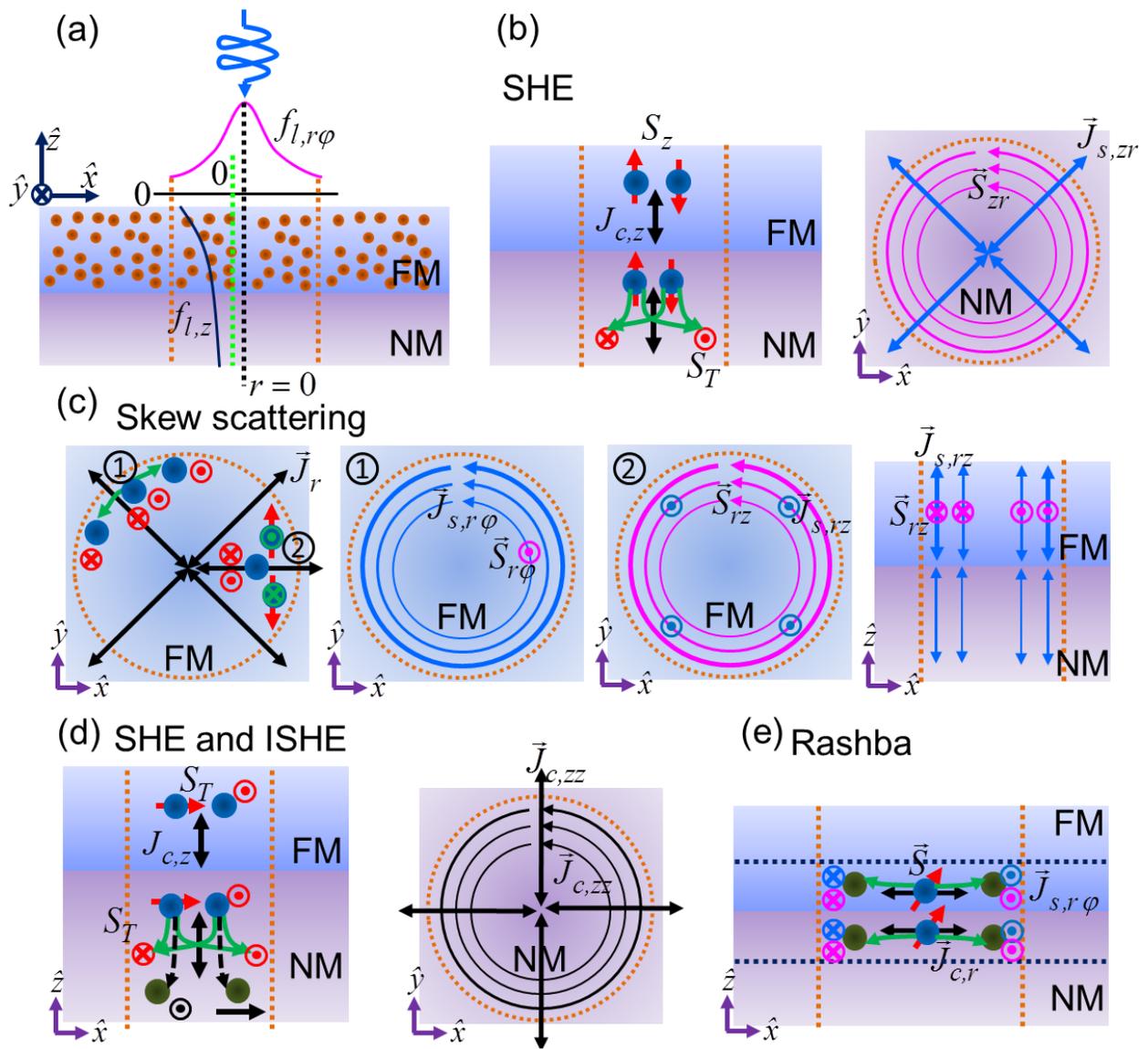

Fig.1

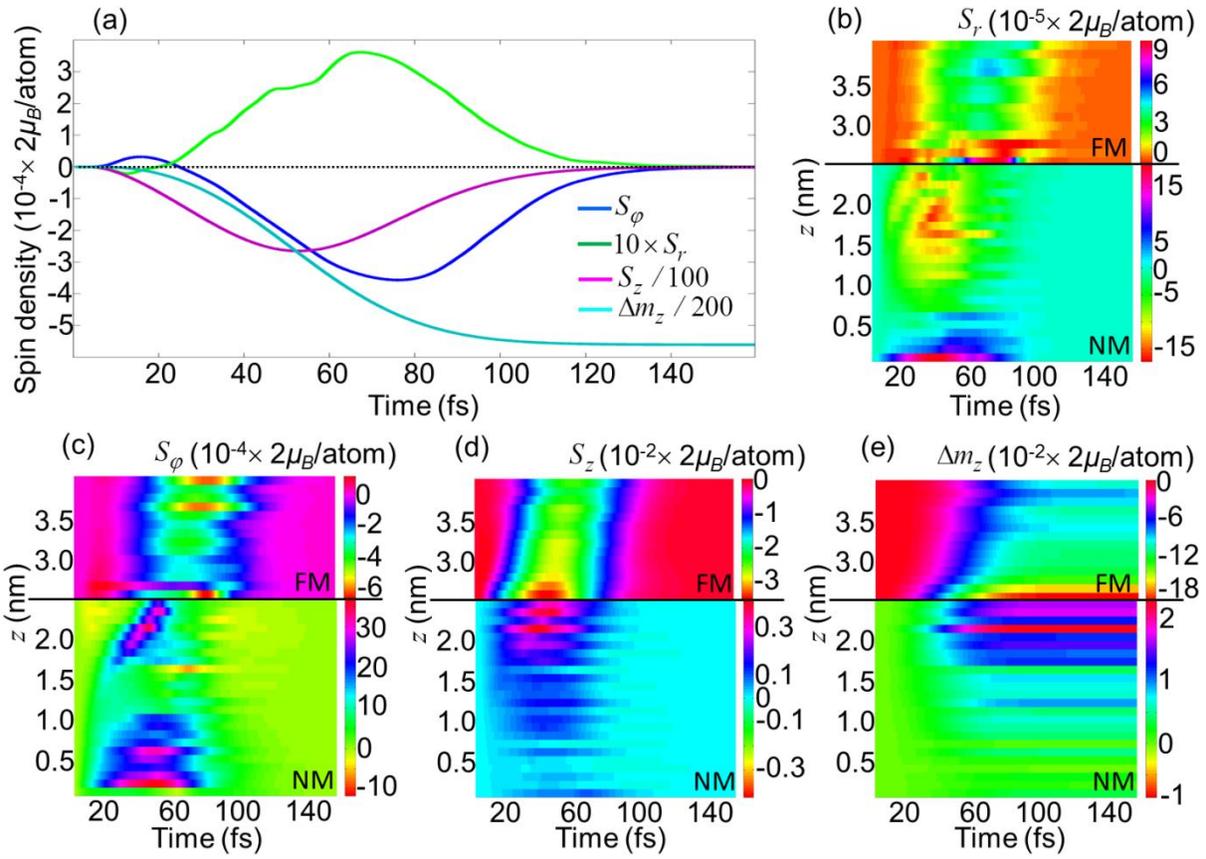

**Fig. 2**

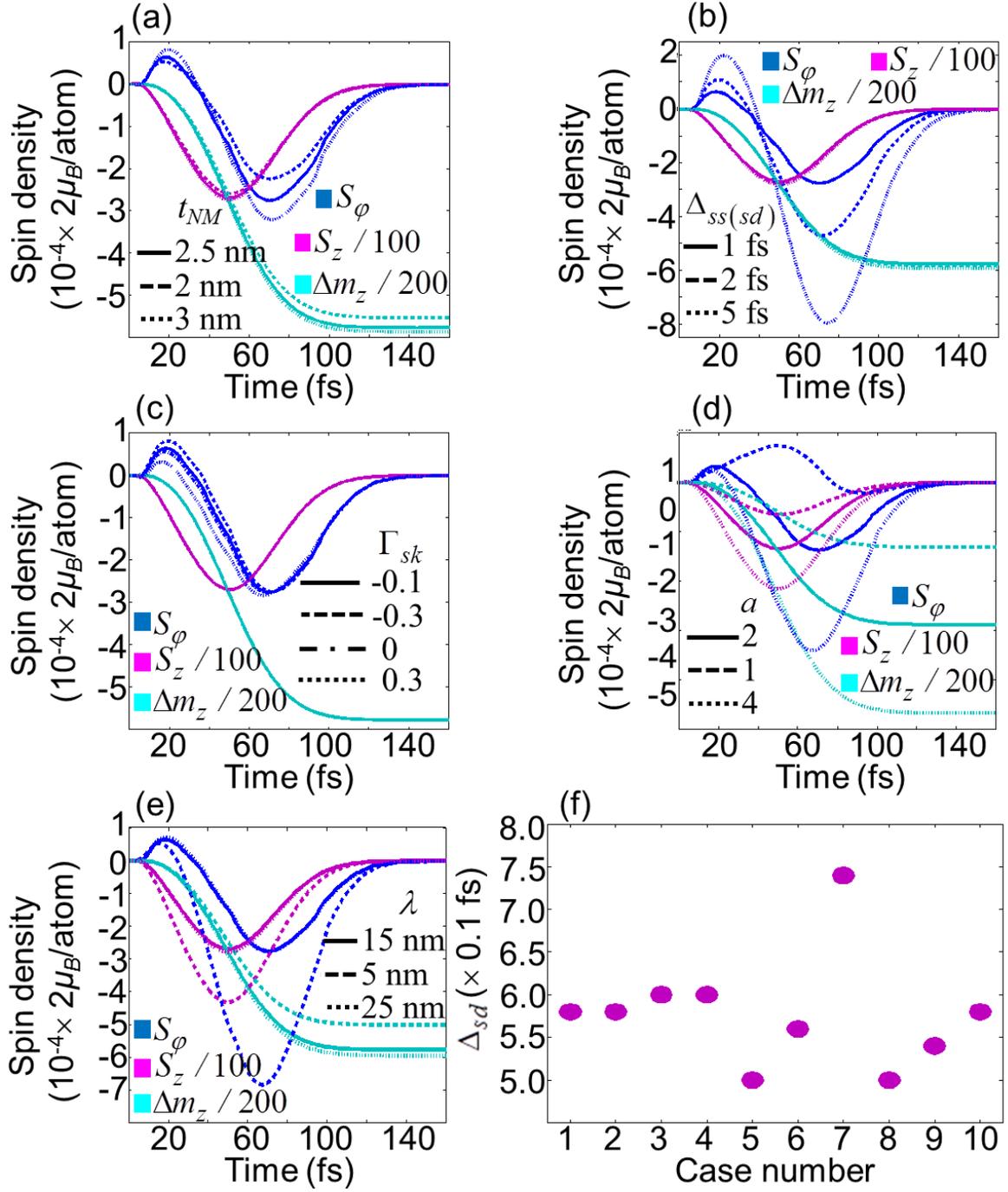

**Fig. 3**

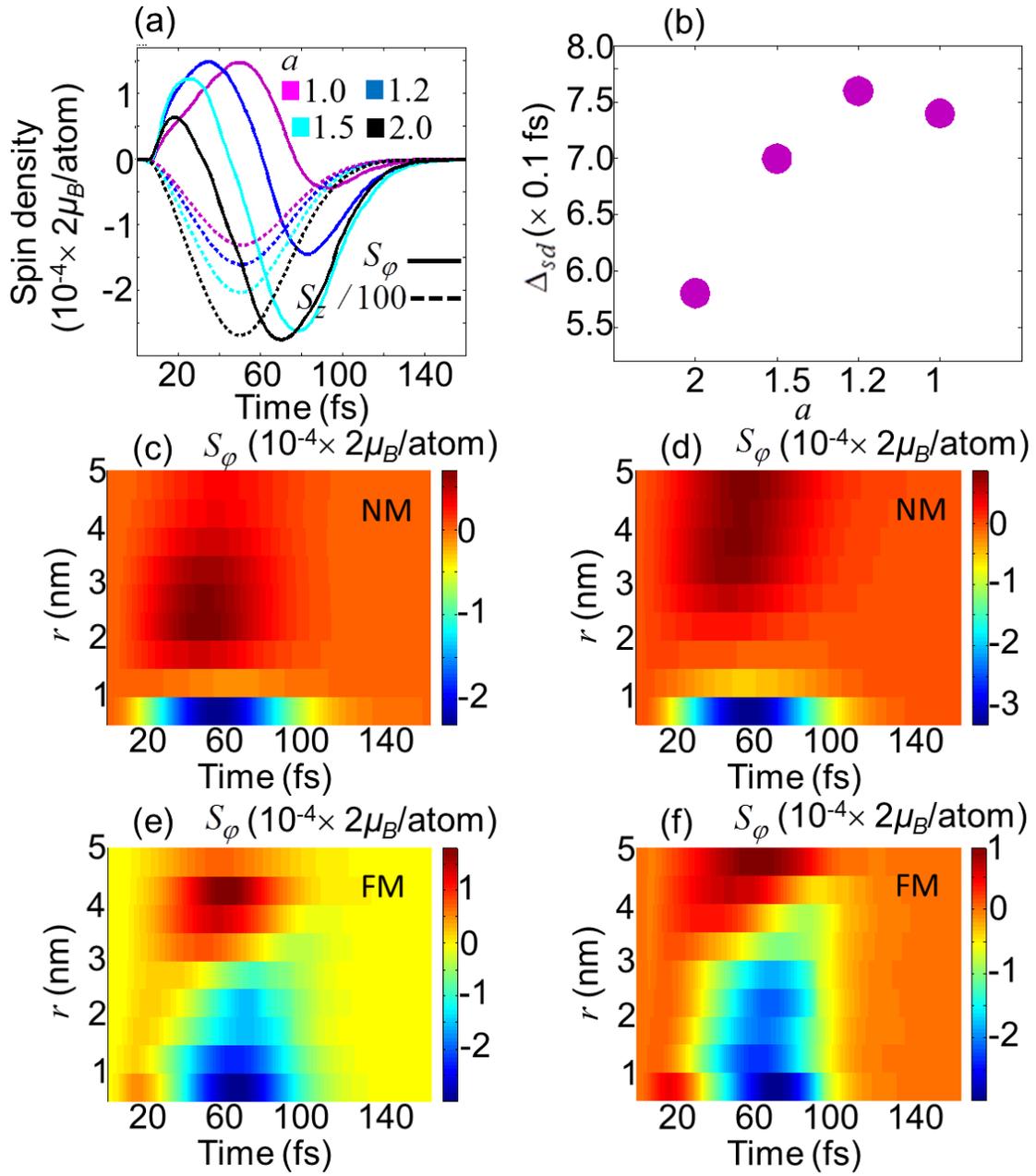

**Fig. 4**